\documentclass[graphicx, 12pt]{article}
\usepackage{indentfirst}

\usepackage{graphicx}
\usepackage{caption}
\usepackage{subfigure}

\begin{document}

\small
\hoffset=-1truecm
\voffset=-2truecm
\title{\bf The fragmentation instability of a black hole with $f(R)$
global monopole under GUP}

\author{Lingshen Chen\hspace {1cm} Hongbo Cheng\footnote
{E-mail address: hbcheng@ecust.edu.cn}\\
Department of Physics,\\ East China University of Science and
Technology,\\ Shanghai 200237, China\\}

\date{}
\maketitle

\begin{abstract}
Having studied the fragmentation of the black holes containing
$f(R)$ global monopole under the generalized uncertainty principle
(GUP), we show the influences from this kind of monopole, $f(R)$
theory and GUP on the evolution of black holes. We focus on the
possibility that the black hole breaks into two parts by means of
the second law of thermodynamics. We derive the entropies of the
initial black hole and the broken parts while the generalization
of Heisenberg's uncertainty principle is introduced. We find that
the $f(R)$ global monopole black hole keeps stable instead of
splitting without the generalization because the entropy
difference is negative. The fragmentation of the black hole will
happen if the black hole entropies are limited by the GUP and the
considerable deviation from the general relativity leads the case
that the mass of one fragmented black hole is smaller and the
other one's is larger.
\end{abstract}

\vspace{6cm} \hspace{1cm} PACS number(s): 04.70.Dy, 14.80.Hv

\noindent \textbf{I.\hspace{0.4cm}Introduction}

The evolution of the universe has been proceeding while the
temperature decreases. During the vacuum phase transition in the
early stage of the universe, various topological defects such as
domain walls, cosmic strings and monopoles may have generated [1,
2]. These topological objects formed due to a breakdown of local
or global gauge symmetries. A global monopole is a spherically
symmetric topological defect arose in the phase transition of a
system composed by a self-coupling triplet of a scalar field whose
original global $O(3)$ symmetry is spontaneously broken to $U(1)$
[3]. There should exist a kind of massive sources that swallow
global monopoles and the metric of the source has a solid deficit
angle [4]. According to the fact of the accelerated expansion of
the universe, the theory of $f(R)$ gravity as a kind of modified
gravity theory was put forward by Buchdahl [5] and has been
applied to explain the accelerated-inflation problem instead of
adding dark energy or dark matter [6-8]. The $f(R)$ gravity
generalizes the general relativity and certainly the
generalization should appear in the description of the background
around the gravitational objects. With the expansion of the
analytical function of the Ricci scalar under the weak field
approximation, the gravitational field of a global monopole in the
modified gravity theory has been explored [9]. It was also found
that the presence of the parameter because of the modification of
gravity is indispensable in providing stable circular orbits for
particles [10]. The metric outside a massive object with $f(R)$
global monopole was obtained [9, 10]. Certainly the components of
this kind of metric contain the terms subject to the global
monopoles and $f(R)$ issue. The numerous investigations have been
paid for the black holes shown with this type of metric. We
considered the gravitational lensing of the massive $f(R)$ global
monopole in the strong field limit [11, 12]. We also calculate the
thermodynamic quantities of this kind of the black hole to examine
the black hole's stability [13]. We derived the greybody factor
for scalar fields in the Schwarzschild spacetime with $f(R)$
global monopole [14]. The timelike naked singularities of the
black hole were probed [15]. The Hawking radiations of the $f(R)$
global monopole black hole was studied based on the Heisenberg's
uncertainty principle or generalized uncertainty principle
respectively [16, 17]. Recently the absorption and scattering of
black hole with $f(R)$ global monopole have been investigated
[18].

The instability of black holes based on the thermodynamics is of
great concern. As a kind of thermodynamic instability, the
fragmentation instability was evaluated under non-perturbation
[19]. The evolutions of black holes are thought to be driven by
the second law of thermodynamics and whether the evolution could
happen depends on the sign of the entropy difference of the black
holes [19]. For a black hole, we derive its entropy. The isolated
black hole is thought to be split and we also derive the total
entropy of the broken black holes [19]. If the final entropy is
smaller than the initial one, the black hole will exist as a
whole, or the black hole will break into the parts [19]. The
fragmentation scheme has been utilized to explore the final fates
of a serious of black holes such as the rotating anti-de Sitter
black holes [20], black holes with a Gauss-Bonnet term [21] and
charged anti-de Sitter black holes [22], etc..

In the research on black holes, it is impossible to neglect the
gravitational effect, so the terms associated with the Newtonian
constant would be added in the Heisenberg's uncertainty principle
[23-29]. The generalized uncertainty principle (GUP) with the
additional terms can be used to cure the divergence from states
density near the black hole horizon and relate the entropy of
black hole to a minimal length as quantum gravity scale [30, 31].
The influence from the GUP modifying the black hole horizon and
definitely changing the black hole entropy further was considered
in the radiation from black holes [30, 31].

It is significant to analyze the fragmentation instability of a
Schwarzschild black hole with global monopole under GUP within the
frame of $f(R)$ gravity. There must exist the gravitational
sources that contain global monopoles in the spacetime governed by
$f(R)$ theory as mentioned above. At the same time the GUP
corrects the horizons subject to the entropies. In the process of
our research on the entropy of a massive source involving the
global monopoles in the accelerated universe, we should not
neglect the influences from global monopole, $f(R)$ approach and
GUP. To our knowledge, few efforts have been contributed to the
fragmentation of a black hole with $f(R)$ global monopole under
GUP. We are going to investigate the possibility that the $f(R)$
global monopole-contained black hole breaks into two sections by
means of the technique from Ref. [19]. We proceed the same
derivation and calculation on entropies of the same black holes
under the GUP. We compare the entropy of the initial black hole
with that of the final system composed of two fragmented black
holes to know how the black hole entropy changes. The sign of the
netropy difference will decide whether the fragmentation will
happen under the second law of thermodynamics. We wonder how the
$f(R)$ gravity and GUP influence on the fragmentation of the black
holes. The results will appear in the end.

\vspace{0.8cm} \noindent \textbf{II.\hspace{0.4cm}The
fragmentation of a black hole with a $f(R)$ global monopole}

We plan to discuss the entropy of black hole with global monopole
in the $f(R)$ gravity. The spherically symmetric metric of the
source was found [9, 10],

\begin{equation}
ds^{2}=A(r)dt^{2}-B(r)dr^{2}-r^2(d\theta^{2}+\sin^{2}\theta
d\varphi^{2})
\end{equation}

\noindent where

\begin{equation}
A(r)=B^{-1}(r)=1-8\pi G\eta^{2}-\frac{2GM}{r}-\psi_{0}(r)
\end{equation}

\noindent and $G$ is the Newtonian constant. As a monopole
parameter, $\eta$ is of the order $10^{16}GeV$ in a typical grand
unified theory, leading $8\pi G\eta^{2}\approx10^{-5}$ [2-4]. $M$
is mass parameter. The factor $\psi_{0}$ reflects the extension of
the Einstein's general relativity. According to $A(r)=0$ from
metric (1), the smaller root is chosen to be an event horizon,

\begin{eqnarray}
r_{H}=r_{H}(M, \eta^{2}, \psi_{0})\hspace{4cm}\nonumber\\
=\frac{(1-8\pi G\eta^{2})-\sqrt{(1-8\pi
G\eta^{2})^{2}-8GM\psi_{0}}}{2\psi_{0}}
\end{eqnarray}

\noindent and the larger one is thought as a cosmological horizon
like,

\begin{equation}
r_{C}=\frac{(1-8\pi G\eta^{2})+\sqrt{(1-8\pi
G\eta^{2})^{2}-8GM\psi_{0}}}{2\psi_{0}}
\end{equation}

\noindent It is obvious that the cosmological horizon will
disappears if the modified parameter $\psi_{0}$ is equal to be
zero. According to Ref. [32-34], the Bekenstein-Hawking entropy of
black hole is proportional to the horizon area,

\begin{equation}
S=\frac{1}{4}A_{H}
\end{equation}

\noindent where

\begin{equation}
A_{H}=4\pi r_{H}^{2}
\end{equation}

The thermodynamic argument for the fragmentation of black hole
claimed that the black hole entropy must increase during the
process in view of the second law of thermodynamics [19]. Here we
assume that the black hole with $f(R)$ global monopole splits into
two parts with the same kind of monopole. Within the
fragmentation, the original black hole can be thought as initial
state and the final state consists of two black holes under the
conservation of mass. Before the splitting, the entropy of the
initial state can be obtained from Eq. (5),

\begin{equation}
S_{i}=\pi r_{H}^{2}(M, \eta^{2}, \psi_{0})
\end{equation}

\noindent After fragmentation, the entropy for the final state is,

\begin{equation}
S_{f}=\pi r_{H}^{2}(\varepsilon_{M}M, \eta^{2}, \psi_{0})+\pi
r_{H}^{2}((1-\varepsilon_{M})M, \eta^{2}, \psi_{0})
\end{equation}

\noindent where $r_{H}(\varepsilon_{M}M, \eta^{2}, \psi_{0})$ and
$r_{H}((1-\varepsilon_{M})M, \eta^{2}, \psi_{0})$ are horizons of
the fragmented black holes with masses $\varepsilon_{M}M$ and
$(1-\varepsilon_{M})M$ respectively from Eq. (3). Following the
same procedure of Ref. [22], we define the mass ratio and its
region $0\leq\varepsilon_{M}\leq1$. We obtain the entropy
difference by substituting the event horizon (3) into the
entropies (7) and (8) with various masses like $\varepsilon_{M}M$
and $(1-\varepsilon_{M})M$,

\begin{eqnarray}
\triangle S=S_{f}-S_{i}\hspace{7cm}\nonumber\\
=\pi\{[\frac{(1-8\pi G\eta^{2})-\sqrt{(1-8\pi
G\eta^{2})^{2}-8\varepsilon_{M}GM\psi_{0}}}{2\psi_{0}}]^{2}\hspace{1cm}\nonumber\\
+[\frac{(1-8\pi G\eta^{2})-\sqrt{(1-8\pi
G\eta^{2})^{2}-8(1-\varepsilon_{M})GM\psi_{0}}}{2\psi_{0}}]^{2}\nonumber\\
-[\frac{(1-8\pi G\eta^{2})-\sqrt{(1-8\pi
G\eta^{2})^{2}-8GM\psi_{0}}}{2\psi_{0}}]^{2}\}\hspace{1cm}
\end{eqnarray}

\noindent If the deviation from the general relativity is ignored,
the difference will recover to be.

\begin{equation}
\lim_{\psi_{0}\longrightarrow 0}\triangle
S=-2\pi\frac{(2GM)^{2}}{(1-8\pi
G\eta^{2})^{2}}\varepsilon_{M}(1-\varepsilon_{M})<0
\end{equation}

\noindent The division of the isolated black hole can not happen
spontaneously because this process makes the entropy decrease
based on the second law of thermodynamics [19]. The negative
nature of the difference above shows that the black hole are
stable instead of rupturing in the case of general relativity
although the weaker influence from global monopole appears. The
influence from the term $8\pi G\eta^{2}$ relating to the monopole
can be neglected because its order is infinitely small. In fact
the existence of the term $8\pi G\eta^{2}$ can not change the sign
of the difference.

As functions of mass ratio $\varepsilon_{M}$ with positive
$\psi_{0}$, the shapes of entropy difference subject to the
generalization of general relativity are plotted in Figure 1. The
greater the magnitude of $\psi_{0}$ is, the larger the absolute
value of the difference is. The $\triangle S$ expressed in Eq. (9)
remains negative with $\psi_{0}>0$, which means that the black
holes keep stable instead of breaking up within the frame of
generalized general relativity no matter whether the sources
contain the global monopoles.

\vspace{0.8cm} \noindent \textbf{III.\hspace{0.4cm}The
fragmentation of black hole with $f(R)$ global monopole under the
generalized uncertainty principle}

The discussion on the fragmentation of a black hole with $f(R)$
global monopole is performed in the context of GUP. The
Heisenberg's uncertainty principle can be generalized within the
microphysics regime as [25, 29-31, 35-44]

\begin{equation}
\bigtriangleup x\bigtriangleup p\geq\frac{\hbar}{2}[1-\frac{\alpha
l_{p}}{\hbar}\bigtriangleup p+(\frac{\beta
l_{p}}{\hbar})^{2}\bigtriangleup p^{2}]
\end{equation}

\noindent and

\begin{equation}
y_{-}\leq y\leq y_{+}
\end{equation}

\noindent where

\begin{eqnarray}
y_{\pm}=(\frac{l_{p}}{\hbar}\bigtriangleup p)_{\pm}\hspace{6.5cm}\nonumber\\
=\frac{1}{2\beta^{2}}(\alpha+\frac{2\bigtriangleup x}{l_{p}})
\pm\frac{1}{2\beta^{2}}(\alpha+\frac{2\bigtriangleup
x}{l_{p}})\sqrt{1-(\frac{2\beta}{\alpha+\frac{2\bigtriangleup
x}{l_{p}}})^{2}}
\end{eqnarray}

\noindent The subscript $\pm$ corresponds to the sign $\pm$ of the
term in the second line in Eq. (13). Here $\alpha$ and $\beta$ are
dimensionless positive parameters modifying the Heisenberg's
inequality. The Planck length is expressed as
$l_{p}=\sqrt{\frac{\hbar G}{c^{2}}}$ with velocity of light in the
vacuum $c$. The terms with the Planck length having something to
do with the Newtonian constant $G$ provide the inequality with the
gravitational effects. Following the procedure of Ref. [30, 31,
45], we choose,

\begin{eqnarray}
\triangle p'=\frac{\hbar}{l_{p}}y_{-}\hspace{7.5cm}\nonumber\\
=\frac{\hbar}{l_{p}}[\frac{1}{2\beta^{2}}(\alpha+\frac{2\bigtriangleup
x}{l_{p}}) -\frac{1}{2\beta^{2}}(\alpha+\frac{2\bigtriangleup
x}{l_{p}})\sqrt{1-(\frac{2\beta}{\alpha+\frac{2\bigtriangleup
x}{l_{p}}})^{2}}]
\end{eqnarray}

\noindent From Eq. (14), the approximate expression for the
uncertainty in the momentum is [45],

\begin{equation}
\triangle p'\approx\frac{\hbar}{\alpha l_{p}+2\triangle x}
\end{equation}

\noindent According to the combination of the GUP and the
approximation (15), we estimate the black hole horizon amended by
the gravitation with the original horizon of black hole as the
lower bound on the region like $\triangle x=2r_{H}$ [30, 31, 45],

\begin{equation}
r'_{H}=r_{H}[1+\frac{(\beta l_{p})^{2}}{4r_{H}(\alpha
l_{p}+4r_{H})}]
\end{equation}

\noindent From expression (16) the correction to the horizon
includes $\alpha$ and $\beta$. It is interesting that no
gravitational effect will act on the black hole horizon with the
disappearance of the $\beta$-term in the GUP no matter whether the
other term with $\alpha$ exists or not. The nonvanishing $\beta$
can keep the correction. The term $(\beta l_{p})^{2}$ determines
the net effect from the correction, so we can discuss the black
hole entropy with the different values of $\beta$ while set
$\alpha$ to be a definite magnitude. Under GUP, the corrected
horizons of black holes certainly lead the corrected entropy
difference,

\begin{equation}
\triangle S'=S'_{f}-S'_{i}
\end{equation}

\noindent where

\begin{equation}
S'_{i}=\pi r'^{2}_{H}(M, \eta^{2}, \psi_{0})
\end{equation}

\begin{equation}
S'_{f}=\pi r'^{2}_{H}(\varepsilon_{M}M, \eta^{2}, \psi_{0})+\pi
r'^{2}_{H}((1-\varepsilon_{M})M, \eta^{2}, \psi_{0})
\end{equation}

\noindent We can hire the Eq. (16) to obtain the corrected radii
$r'_{H}(M, \eta^{2}, \psi_{0})$, $r'_{H}(\varepsilon_{M}M,
\eta^{2}, \psi_{0})$ and $r'_{H}((1-\varepsilon_{M})M, \eta^{2},
\psi_{0})$ as follow,

\begin{equation}
r'_{H}(M, \eta^{2}, \psi_{0})=r_{H}(M, \eta^{2},
\psi_{0})+\frac{(\beta l_{p})^{2}}{4(\alpha l_{p})+16r_{H}(M,
\eta^{2}, \psi_{0})}
\end{equation}

\begin{equation}
r'_{H}(\varepsilon_{M}M, \eta^{2}, \psi_{0})=r'_{H}(M, \eta^{2},
\psi_{0})|_{M\longrightarrow\varepsilon_{M}M}
\end{equation}

\begin{equation}
r'_{H}((1-\varepsilon_{M})M, \eta^{2}, \psi_{0})=r'_{H}(M,
\eta^{2}, \psi_{0})|_{M\longrightarrow(1-\varepsilon_{M})M}
\end{equation}

\noindent Here $r'_{H}(M, \eta^{2}, \psi_{0})$ stands for the
GUP-limited horizon of the initial black hole and
$r'_{H}(\varepsilon_{M}M, \eta^{2}, \psi_{0})$ and
$r'_{H}((1-\varepsilon_{M})M, \eta^{2}, \psi_{0})$ are the horizon
radii of the separated black holes belonging to the final state
respectively under GUP.

We should consider the fragmentation of an $f(R)$ global monopole
black hole under the GUP and that the entropy of an isolated
system can not decrease in any spontaneous process [19, 46]. The
nature of the entropy difference during the evolution of the black
hole helps us to determine whether the black hole will split.
According to that the monopoles probably formed in the process of
the vacuum phase transition early in the universe and the fact
that the universe expands faster mentioned above, the
gravitational sources involving $f(R)$ global monopoles exist
inevitably. It should be pointed out that some kinds of black
holes are the special cases of the gravitational sources. The
black holes limited by the GUP because of the requirement of a
minimal length of the order of Planck length. At first we ignore
the correction to the general relativity and show the entropy
difference as a function of ratio $\varepsilon_{M}$ graphically in
Figure 2. It is found that there may be two roots
$\varepsilon_{M0}$ and $1-\varepsilon_{M0}$ from $\triangle S'=0$.
If $\varepsilon_{M}<\varepsilon_{M0}$ or
$\varepsilon_{M}>1-\varepsilon_{M0}$, then $\triangle S'>0$, which
means that the black hole will break into two parts under the GUP.
The farther the inequality deviates from the Heisenberg's
uncertainty principle, the larger the value of $\triangle S'$ is.
The Figure 3 indicates how the generalization of general
relativity affects the entropy difference $\triangle S'$ within
the frame of GUP. The generalization will adjust the distribution
of the masses for the portions of black hole. The larger
$\psi_{0}$ makes $\varepsilon_{M0}$ smaller, meaning that the
black hole is divided into two sources, tiny one and the other
huge in their masses. We sum up the results from the Figure 2 and
Figure 3 to exhibit the relation among the $\varepsilon_{N0}$,
$\psi_{0}$ and $\beta$ in Figure 4. The appearance of GUP must
lead the black holes swallowing $f(R)$ global monopoles to be
divided. The weaker $\beta$-term in GUP and the stronger
modification on the standard general relativity cause the black
hole to become one part with extremely small mass and the other
one with extremely large mass. Here we fix the value of $\alpha$
for simplicity during our discussion.

\vspace{0.8cm} \noindent \textbf{IV.\hspace{0.4cm}Conclusion}

We discuss the fragmentation instability of black hole with $f(R)$
global monopole under the GUP. The influence that the global
monopoles bring about is too small to be considered. At first we
employ the approach of the non-perturbative fragmentation from
Ref. [19] to compare the entropy of the initial $f(R)$ global
monopole black hole limited by the standard Heisenberg's
inequality with the sum of those of two fragmented ones. According
to the second law of thermodynamics, only the fragmentation of
black hole with the nonnegative nature of the entropy difference
can happen. The sign of the entropy difference depends on whether
the traditional principle of indeterminacy has been revised. Under
the original indeterminate principle, the nature of the entropy
difference will remain negative no matter whether the general
relativity has been modified. When the GUP is introduced, the
entropy of initial black hole will be smaller than that of the
final system consisting of two fragmented black holes. The more
manifest the deviation from the Heisenberg's certainty principle
is, the smaller the difference of the masses for the two
fragmented black holes is. In this case the more considerable
correction of $f(R)$ just regulate the division of the mass of
initial black hole, resulting in that one fragmented black hole
mass is tiny and other is huge. Our results show the influences
from the global monopole, modified general relativity and
generalized uncertainty principle on Schwarzschild black hole
fragmentation.

\vspace{1cm}
\noindent \textbf{Acknowledge}

This work is supported by NSFC No. 10875043.

\newpage
\begin{figure}
\setlength{\belowcaptionskip}{10pt} \centering
\includegraphics[width=15cm]{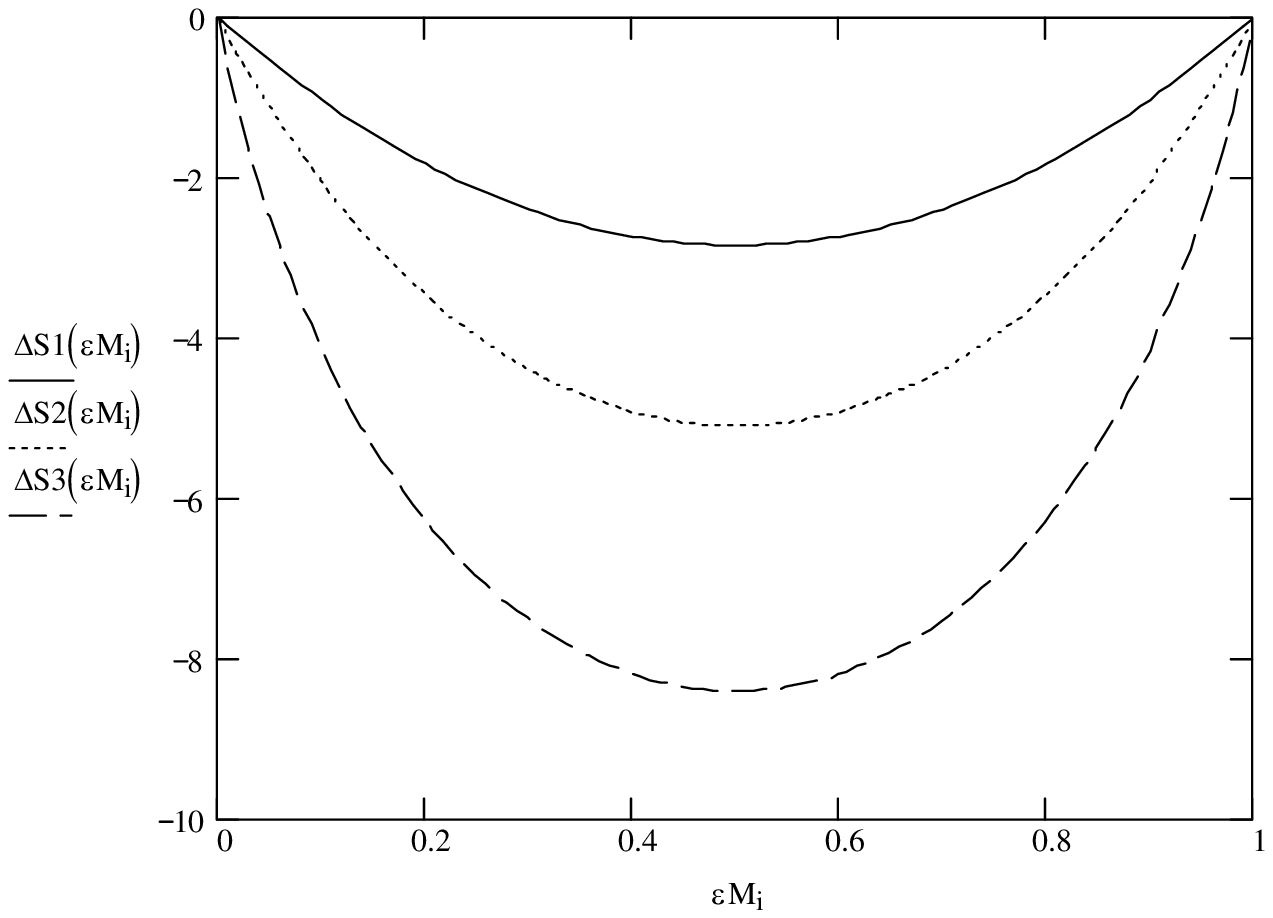}
\caption{The solid, dotted, dashed curves of the dependence of
entropy difference $\triangle S$ on $\varepsilon_{M}$ in unit of
$\pi(GM)^{2}$ for $\psi_{0}=0.05, 0.1, 0.12$ respectively without
GUP.}
\end{figure}

\begin{figure}
\setlength{\belowcaptionskip}{10pt} \centering
\includegraphics[width=15cm]{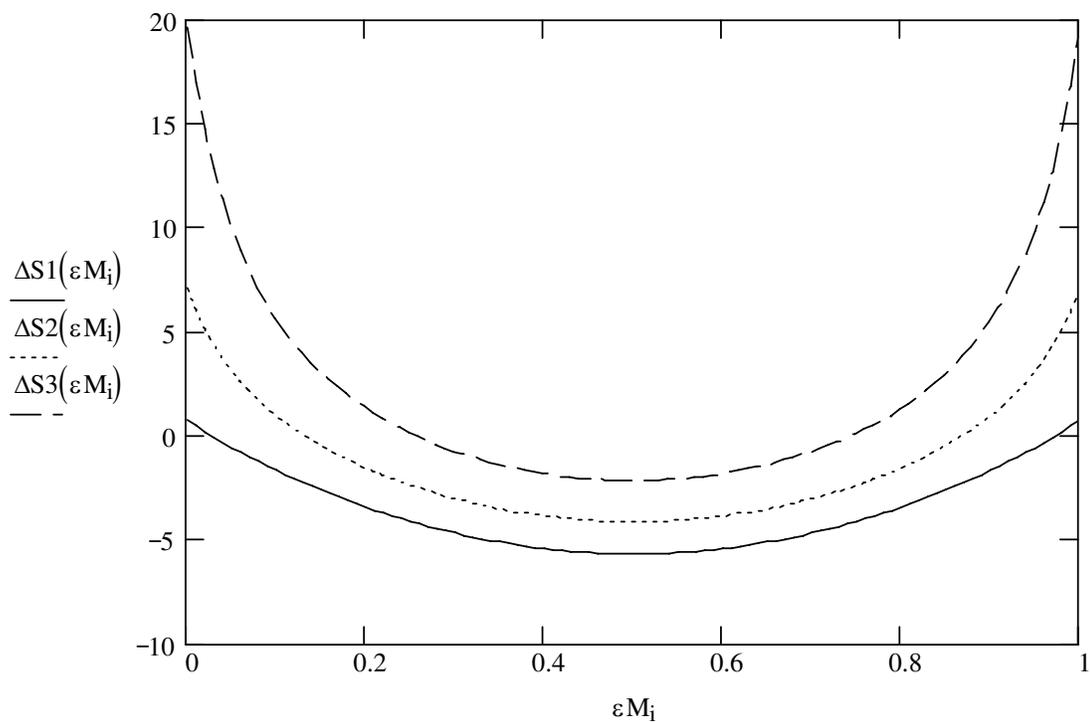}
\caption{The solid, dotted, dashed curves of the dependence of
entropy difference $\triangle S'$ on $\varepsilon_{M}$ in unit of
$\pi(GM)^{2}$ for $(\beta l_{p})^{2}=2, 6, 10$ respectively with
$\psi_{0}=0$ and $\alpha l_{p}=1$ for simplicity.}
\end{figure}

\newpage
\begin{figure}
\setlength{\belowcaptionskip}{10pt} \centering
\includegraphics[width=15cm]{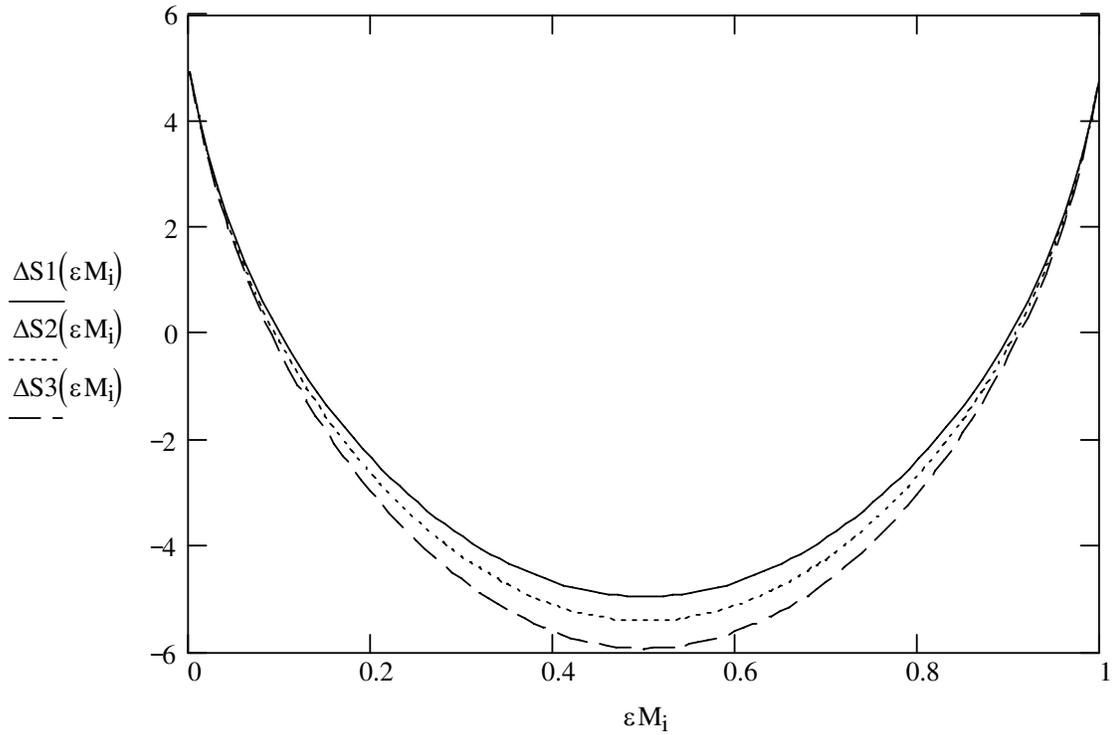}
\caption{The solid, dotted, dashed curves of the dependence of
entropy difference $\triangle S'$ on $\varepsilon_{M}$ in unit of
$\pi(GM)^{2}$ for $\psi_{0}=0.03, 0.02, 0.01$ respectively with
$(\beta l_{p})^{2}=5$ and $\alpha l_{p}=1$ for simplicity.}
\end{figure}

\newpage
\begin{figure}
\setlength{\belowcaptionskip}{10pt} \centering
\includegraphics[width=15cm]{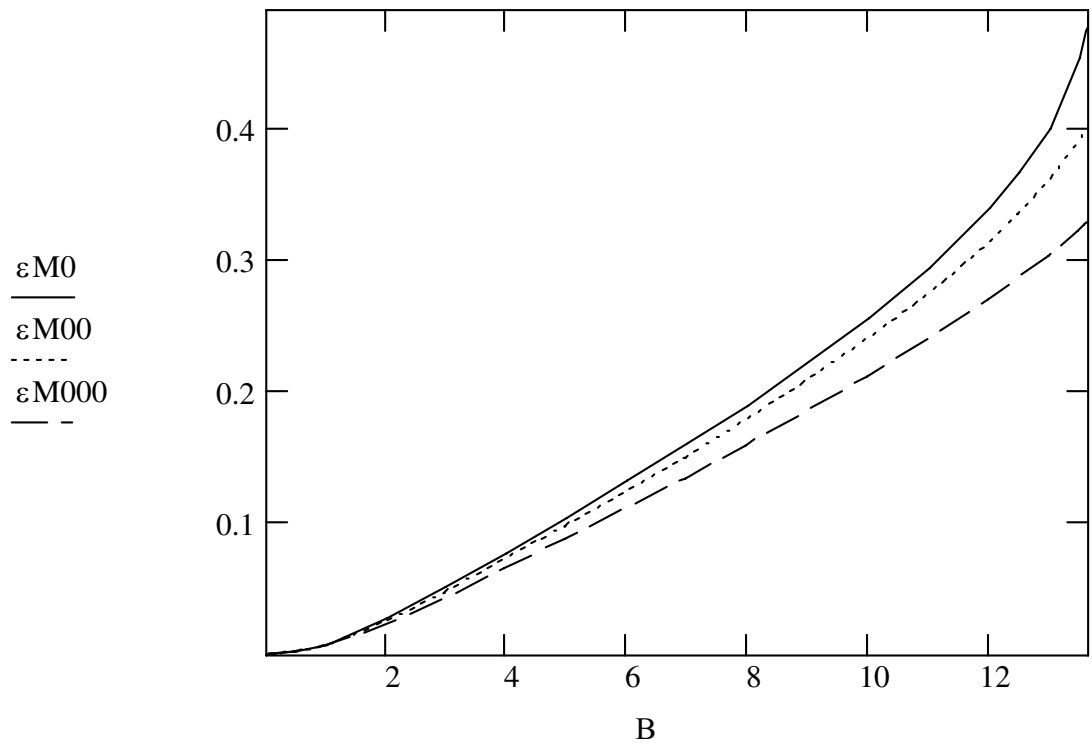}
\caption{The solid, dashed, dot curves of the relations between
$\varepsilon_{M0}$ and $B=l_{p}\beta$ for $\psi_{0}=0.01, 0.02,
0.03$ respectively while $\alpha l_{p}=1$ for simplicity.}
\end{figure}

\end{document}